\documentclass[aps,prd,11pt,
nofootinbib,superscriptaddress]{revtex4-2}

\usepackage{amssymb, amsfonts, amsmath, bm}
\usepackage[
]{graphicx}
\usepackage{color}
\usepackage{mathrsfs}

\usepackage[
]{hyperref}
\hypersetup{%
 colorlinks=true,
  linkcolor=blue,   
  citecolor=blue,   
  urlcolor=blue   
}
\newcommand{\orcid}[1]{%
  \href{https://orcid.org/#1}{%
   \IfFileExists{orcid.pdf}{\includegraphics[height=0.7em]{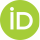}}{}%
 }%
}

\begin{document}
\title{%
Strong-deflection expansion of the deflection angle near\\
a degenerate photon sphere
}
\author{
Takahisa Igata\:\!\orcid{0000-0002-3344-9045}
}
\email{takahisa.igata@gakushuin.ac.jp}
\affiliation{
Department of Physics, Faculty of Science, Gakushuin University, Tokyo 171-8588, Japan
}
\author{Tadashi Sasaki\:\!\orcid{0000-0002-2172-8653}}
\email{ta-sasaki@kumagaku.ac.jp}
\affiliation{Department of Commerce, Faculty of Commerce, Kumamoto Gakuen University, Kumamoto 862-8680, Japan}
\author{Naoki Tsukamoto\:\!\orcid{0000-0003-4871-5155}}
\email{tsukamoto@rikkyo.ac.jp}
\affiliation{Department of Physics, Faculty of Science,
Tokyo University of Science, Tokyo 162-8601, Japan}

\date{\today}

\begin{abstract}
We present a strong-deflection expansion for the deflection angle of light rays scattered near a degenerate photon sphere in asymptotically flat, static, and spherically symmetric spacetimes. Our prescription isolates the divergent contribution to the deflection-angle integral arising from the ray's passage near the marginal orbit in a way that remains nonsingular at marginality, thereby yielding a unique leading power-law term. When expressed in terms of the radius of closest approach, the leading coefficient in the strong deflection limit factorizes into a universal branch constant and a local factor determined by the third derivative of the effective potential at the degenerate photon sphere. When the expansion is rewritten in terms of the impact parameter, the coefficient is simply multiplied by an additional local conversion factor. We show that the local factor in the closest-approach expansion admits an invariant representation through the areal-radius derivative of a dimensionless tidal measure constructed from the electric part of the Weyl tensor. In general relativity, we further relate this quantity to the areal-radius derivative of a weighted null-energy density profile. Analytic examples validate this factorization and yield closed-form expressions for the leading divergent coefficients in representative marginal configurations.
\end{abstract}

\maketitle

\section{Introduction}
\label{sec:1}
Gravitational light deflection in the strong-field regime is governed by unstable circular photon orbits, which delineate the shadow boundary and generate ring-like image features~\cite{Bardeen:1973}. The Event Horizon Telescope observations of M87* and Sgr A*~\cite{EventHorizonTelescope:2019dse,EventHorizonTelescope:2022wkp} have renewed interest in these phenomena, motivating accurate theoretical modeling of the shadow boundary~\cite{Luminet:1979,Fukue:1988,Falcke:1999pj,Takahashi:2004xh,Hioki:2009na} and of photon-ring and lensing-ring structures~\cite{Igata:2019pgb,Gralla:2019xty,Gralla:2020srx}.

In asymptotically flat spacetimes with a generic (nondegenerate) unstable circular photon orbit, the deflection angle $\hat{\alpha}$ of scattering null geodesics diverges logarithmically in the strong deflection limit (SDL) as the impact parameter $b$ approaches its critical value $b_{\mathrm{c}}$ (e.g., in the Schwarzschild case~\cite{Darwin:1959,Bozza:2001xd}). In gravitational lensing, this logarithmic divergence produces an infinite sequence of relativistic images~\cite{Virbhadra:1999nm,Perlick:2004}. Their observables can be expressed in terms of a small set of SDL coefficients, which determine the leading logarithmic term and the leading finite offset in the strong-deflection expansion of $\hat{\alpha}$~\cite{Bozza:2002zj,Bozza:2010xqn}. Because the leading logarithmic coefficient is determined locally by the geometry at the unstable circular photon orbit, strong-lensing observables provide direct probes of the strong-field region.

In spherically symmetric spacetimes, circular photon orbits form a photon surface in the sense that any null geodesic initially tangent to the surface remains tangent to it~\cite{Claudel:2000yi}. We refer to this photon surface as a photon sphere for radially unstable circular photon orbits and as an anti-photon sphere for radially stable orbits~\cite{Cvetic:2016bxi}; only the unstable case generically leads to a divergent deflection angle, whereas stable circular photon orbits can yield a finite deflection angle~\cite{Kudo:2022ewn}. At critical values of the spacetime parameters, a photon sphere and an anti-photon sphere can coalesce into a degenerate photon sphere~\cite{Hod:2017zpi,Cunha:2017qtt}, whose circular photon orbits are marginally unstable. For wormholes, such a coalescence may involve not only photon spheres and anti-photon spheres but also throat-related structures~\cite{Zhang:2024sgs,Tsukamoto:2024pid}. In these degenerate cases, the SDL behavior of the deflection angle changes qualitatively: $\hat{\alpha}$ exhibits a power-law divergence rather than the usual logarithmic divergence~\cite{Tsukamoto:2020iez}. The power-law divergence in the SDL has been analyzed in many spacetimes, including Reissner--Nordstr\"om naked singularities~\cite{Tsukamoto:2025hbz}, Hayward regular black holes~\cite{Chiba:2017nml,Tsukamoto:2025hbz}, Damour--Solodukhin wormholes~\cite{Damour:2007ap,Tsukamoto:2020uay}, an effective loop quantum gravity black hole at a critical parameter value~\cite{Fu:2021fxn}, and Majumdar--Papapetrou di-holes~\cite{Patil:2016oav}.

For static and spherically symmetric spacetimes, Ref.~\cite{Tsukamoto:2020iez} developed a general strong-deflection expansion for scattering near a degenerate photon sphere and predicted $\hat{\alpha} \propto |b/b_{\mathrm{c}} - 1|^{-1/6}$. Subsequent studies of the marginal Reissner--Nordstr\"om naked singularity using a Picard--Fuchs approach~\cite{Sasaki:2025web} and an independent method~\cite{Tsukamoto:2025hbz} (following Ref.~\cite{Eiroa:2002mk}) confirmed the robustness of the $-1/6$ scaling, while showing that the leading coefficient must be extracted using a prescription that remains nonsingular at marginality.

At marginality, the standard SDL prescription---isolating the logarithmically divergent near-orbit contribution to the deflection-angle integral (see, e.g., Refs.~\cite{Bozza:2002zj,Tsukamoto:2016jzh,Igata:2025taz})---becomes singular because the quadratic instability of the unstable circular photon orbit vanishes. The divergence is then governed by the first nonvanishing derivative of the effective potential beyond quadratic order, so a different prescription for extracting the near-critical contribution is needed to obtain unique leading coefficients.

By contrast, for nondegenerate photon spheres, recent progress has shown that the leading logarithmic SDL coefficient admits a manifestly coordinate-invariant expression in terms of local curvature (and, in general relativity, matter variables) evaluated at the photon sphere~\cite{Igata:2025taz}. This invariant viewpoint also clarifies the dynamics: the logarithmic divergence rate is tied to the instability (Lyapunov) exponent of the circular photon orbit~\cite{Stefanov:2010xz,Raffaelli:2014ola}, which can be derived covariantly from the geodesic-deviation equation and interpreted as the growth rate of transverse deviations driven by the local tidal field~\cite{Igata:2026hzb} (see also related extensions to axisymmetric settings~\cite{Igata:2025plb,Igata:2025hpy}).

Motivated by recent covariant results for the nondegenerate SDL, we formulate a strong-deflection expansion whose leading coefficients are determined by local, coordinate-invariant data. Specifically, we ask which local invariants govern the power-law divergence near a degenerate photon sphere and---within general relativity---how they encode the local stress-energy distribution.

In this paper, we develop a systematic strong-deflection expansion for the deflection angle of light rays scattered near a degenerate photon sphere in general static and spherically symmetric spacetimes. We isolate the divergent contribution to the deflection-angle integral in a manner that remains well defined at marginality and thereby obtain a canonical near-critical expansion in terms of the areal radius of closest approach $R_0$ of the form $\hat{\alpha}\simeq c_s |R_0/R_{\mathrm{c}}-1|^{-1/2}$, where $R_{\mathrm{c}}$ is the areal radius of the marginally unstable circular photon orbit. We show that $c_s$ factorizes into a branch-dependent universal constant and a local marginal-instability factor determined by the first nonvanishing derivative---here, the third derivative---of the effective potential at the marginally unstable circular photon orbit. Rewriting the result in terms of the impact parameter $b$, we then obtain $\hat{\alpha}\simeq \bar{c}_s |b/b_{\mathrm{c}}-1|^{-1/6}$, with $\bar{c}_s$ related to $c_s$ by a local conversion factor. We further express the relevant marginal-instability data in terms of locally measurable curvature in an orthonormal tetrad, identifying the key quantity with the areal-radius derivative of a dimensionless tidal-field measure built from the electric part of the Weyl tensor. Specializing to general relativity, we relate this characterization to local matter variables, in particular, to the null-energy density along the photon-orbit direction and to the areal-radius derivative of a suitably weighted null-energy-density profile near the marginal orbit.

This paper is organized as follows. In Sec.~\ref{sec:2}, we set up the deflection angle for scattering null geodesics in a general static and spherically symmetric spacetime. In Sec.~\ref{sec:3}, we characterize marginally unstable circular photon orbits and the associated critical trajectory. In Sec.~\ref{sec:4}, we derive the power-law behavior from a near-critical expansion of the deflection-angle integral and obtain the leading coefficients and the finite part. In Sec.~\ref{sec:5}, we rewrite the marginal-instability conditions and the SDL coefficients first in terms of local curvature and then, within general relativity, in terms of the energy density and principal pressures. In Sec.~\ref{sec:6}, we illustrate the formalism with analytic examples, and in Sec.~\ref{sec:7}, we summarize our results. Throughout this paper, we use geometrized units with $G=1$ and $c=1$ and, when convenient, employ abstract index notation~\cite{Wald:1984}.

\section{Deflection angle in static and spherically symmetric spacetimes}
\label{sec:2}
We consider a static and spherically symmetric spacetime with the line element 
\begin{align}
\mathrm{d}s^2=-A(r)\:\!\mathrm{d}t^2+B(r)\:\!\mathrm{d}r^2+R(r)^2 (\mathrm{d}\theta^2+\sin^2\theta\:\!\mathrm{d}\varphi^2),
\label{eq:metric}
\end{align}
where $A(r)$ and $B(r)$ are positive in the domain of interest and $R(r)$ is the areal radius. Here and throughout the paper, a prime denotes differentiation with respect to $r$. On any interval where $R'(r)>0$, one may equivalently choose the areal-radius gauge $R=r$. We nevertheless keep a generic monotonic radial coordinate $r$ rather than fixing this gauge from the outset. This allows us to formulate the analysis in a way that avoids an early gauge choice and to express the final results in terms of geometrically meaningful quantities built from the areal radius. We assume asymptotic flatness, $A(r)\to 1$, $B(r)\to 1$, and $R(r)\to r$, as $r\to \infty$.

We also introduce the Misner--Sharp mass~\cite{Misner:1964je,Hayward:1994bu,Kinoshita:2024wyr}, defined by
\begin{align}
m(r)
&\equiv \frac{R(r)}{2}\left[\:\!
1-g^{ab} (\nabla_a R)(\nabla_b R)
\:\!\right]
\label{eq:MSdef}
\\
&=\frac{R(r)}{2}\left[\:\!1-\frac{(R')^2}{B(r)}\:\!\right],
\label{eq:MS}
\end{align}
where $g^{ab}$ is the inverse of the metric and the second equality follows from Eq.~\eqref{eq:metric}. The Misner--Sharp mass characterizes the deviation of the quasilocal geometry from flatness. In asymptotically flat spacetimes, it approaches the Arnowitt--Deser--Misner (ADM) mass as $r \to \infty$.

To describe the motion of massless test particles, we consider null geodesics. Exploiting spherical symmetry, we restrict attention to the equatorial plane $\theta = \pi/2$ without loss of generality. The geodesic equations follow from the Lagrangian
\begin{align}
\mathscr{L}
=\frac{1}{2}\left[\:\!-A(r)\:\!\dot{t}^{\:\!2}+B(r)\:\!\dot{r}^{\:\!2}+R(r)^2\dot{\varphi}^{\:\!2}\:\!\right],
\end{align}
where an overdot denotes differentiation with respect to an affine parameter $\lambda$. Because the spacetime is static and spherically symmetric, $\mathscr{L}$ is independent of $t$ and $\varphi$. The corresponding constants of motion are the energy $E=A(r)\:\!\dot{t}$ and the angular momentum $L=R(r)^2 \dot{\varphi}$.

For null geodesics, the Lagrangian vanishes, $\mathscr{L}=0$, yielding the constraint
\begin{align}
-A(r)\:\!\dot{t}^{\:\!2}+B(r)\:\!\dot{r}^{\:\!2}+R(r)^2\dot{\varphi}^{\:\!2}=0.
\label{eq:nullcond}
\end{align}
Using the definitions of $E$ and $L$ to eliminate $\dot{t}$ and $\dot{\varphi}$, we obtain the radial equation of motion
\begin{align}
\dot{r}^{\:\!2}+\frac{1}{B(r)}\left[\:\!
\frac{L^2}{R(r)^2}-\frac{E^2}{A(r)}
\:\!\right]=0. 
\end{align}
To describe the spatial trajectory, we eliminate the affine parameter $\lambda$ in favor of the azimuthal angle $\varphi$. Restricting attention to nonradial null geodesics ($L\neq 0$) and using $\dot{r}=(\mathrm{d}r/\mathrm{d}\varphi)\dot{\varphi}$ together with $L=R(r)^2 \dot{\varphi}$, we obtain the equation governing the trajectory 
\begin{align}
\left(
\frac{\mathrm{d}r}{\mathrm{d}\varphi}
\right)^2+V(r)=0,
\label{eq:orbitaleq}
\end{align}
where the effective potential $V(r)$ is defined as
\begin{align}
V(r)=\frac{R(r)^2}{B(r)}\left[\:\!
1-\frac{R(r)^2}{b^2A(r)}
\:\!\right],
\label{eq:V}
\end{align}
where $b\equiv L/E$ is the impact parameter. Equation~\eqref{eq:orbitaleq} implies that physical motion is restricted to regions where $V(r) \le 0$. Since $b=L/E$ fully characterizes the null trajectory, the motion is independent of the photon energy $E$ and depends only on the spacetime geometry and $b$.

We focus on scattering trajectories, in which a photon comes from infinity, reaches a minimum areal radius, and returns to infinity. Such trajectories have a single turning point at the radius of closest approach $r_0$, where the radial velocity vanishes. Accordingly, the turning point satisfies $V(r_0)=0$. Evaluating this condition using Eq.~\eqref{eq:V}, we obtain a direct relation between the impact parameter and the metric functions at the turning point,
\begin{align}
b^2=\frac{R_0^2}{A_0},
\label{eq:bsq}
\end{align}
where $R_0\equiv R(r_0)$ and $A_0\equiv A(r_0)$. Hereafter, the subscript 0 denotes evaluation at $r=r_0$. Substituting Eq.~\eqref{eq:bsq} into Eq.~\eqref{eq:V}, we can rewrite the effective potential as
\begin{align}
V(r)=\frac{R(r)^2}{B(r)}\left(
1-\frac{R(r)^2}{R_0^2}\frac{A_0}{A(r)}
\right).
\end{align}

The total change in the azimuthal angle along such a scattering trajectory is obtained by integrating Eq.~\eqref{eq:orbitaleq} from $r_0$ to infinity. Exploiting the symmetry of the trajectory about the turning point, the total angular variation is 
\begin{align}
\Delta \varphi(R_0)=2\int_{r_0}^\infty \frac{\mathrm{d}r}{\sqrt{-V(r)}}.
\label{eq:totalchange}
\end{align}
Because $R'(r)>0$, specifying $R_0$ is equivalent to specifying $r_0$. The deflection angle $\hat{\alpha}$ is defined as the deviation from the straight-line value $\pi$ in flat spacetime, namely,
\begin{align}
\hat{\alpha}(R_0)=\Delta\varphi(R_0)-\pi.
\label{eq:deflectionangle}
\end{align}
We parametrize scattering trajectories by $R_0$ (equivalently $b$).

\section{Marginally unstable circular photon orbits}
\label{sec:3}
In this section, we consider marginally unstable circular photon orbits, for which the usual quadratic behavior of the effective potential $V(r)$ near the orbit becomes degenerate. The corresponding critical noncircular null geodesic asymptotically approaches the circular orbit and undergoes an arbitrarily large azimuthal advance in its vicinity. This behavior leads to a divergent contribution to the deflection angle in the SDL. 

For a circular photon orbit at $r=r_{\mathrm{c}}$, the radial coordinate remains constant along the geodesic. Accordingly, the effective potential $V(r)$ appearing in Eq.~\eqref{eq:orbitaleq} has a double root at $r=r_{\mathrm{c}}$, namely $V(r_{\mathrm{c}})=0$ and $V'(r_{\mathrm{c}})=0$. Substituting Eq.~\eqref{eq:V} into $V(r_{\mathrm{c}})=0$ yields 
\begin{align}
b^2=
\frac{R_{\mathrm{c}}^2}{A_{\mathrm{c}}}\equiv b_{\mathrm{c}}^{\,2},
\label{eq:bcsq}
\end{align}
where $A_{\mathrm{c}}\equiv A(r_{\mathrm{c}})$ and $R_{\mathrm{c}}\equiv R(r_{\mathrm{c}})$. We denote by $b_{\mathrm{c}}$ the critical impact parameter associated with the circular photon orbit. Hereafter, the subscript $\mathrm{c}$ denotes evaluation at $r=r_{\mathrm{c}}$. Using Eq.~\eqref{eq:bcsq}, the condition $V'_{\mathrm{c}}=0$ implies
\begin{align}
\frac{A'_{\mathrm{c}}}{A_{\mathrm{c}}}-\frac{2R'_{\mathrm{c}}}{R_{\mathrm{c}}}=0, 
\label{eq:ccond}
\end{align}
which provides a local condition---written only in terms of the metric functions and their derivatives---for the existence of a circular photon orbit in a general static and spherically symmetric spacetime.

The stability of a circular photon orbit is determined by the sign of $V''_{\mathrm{c}}$: a circular photon orbit is radially unstable for $V''_{\mathrm{c}}<0$ (a local maximum of $V$) and radially stable for $V''_{\mathrm{c}}>0$ (a local minimum of $V$). In the marginal case $V''_{\mathrm{c}}=0$, the stability is controlled by higher derivatives. Here, we define a marginally unstable circular photon orbit as one satisfying the following conditions (we leave higher-order degeneracies for future work):
\begin{align}
V_{\mathrm{c}}=0, \qquad V'_{\mathrm{c}}=0, \qquad V''_{\mathrm{c}}=0, \qquad V'''_{\mathrm{c}}<0.
\label{eq:mupco}
\end{align}
The explicit expressions for these quantities are given in Eqs.~\eqref{eq:VcH}--\eqref{eq:VppcH} and Eq.~\eqref{eq:V3_H3} in the Appendix. In many spacetimes, this marginal configuration arises when an unstable and a stable circular photon orbit coalesce as a parameter is varied. Using Eqs.~\eqref{eq:bcsq} and~\eqref{eq:ccond}, we can rewrite the marginal condition $V''_{\mathrm{c}}=0$ as
\begin{align}
\frac{R''_{\mathrm{c}}}{R_{\mathrm{c}}}+\left(
\frac{R'_{\mathrm{c}}}{R_{\mathrm{c}}}
\right)^2-
 \frac{A''_{\mathrm{c}}}{2A_{\mathrm{c}}}=0. 
\label{eq:Vppc2}
\end{align}
While Eq.~\eqref{eq:ccond} determines the location of the circular photon orbit, Eq.~\eqref{eq:Vppc2} further constrains the spacetime parameters and identifies the critical configuration at which the orbit becomes marginally unstable.

We focus on the critical noncircular trajectory with $b=b_{\mathrm{c}}$, which asymptotically approaches the marginally unstable circular orbit at $r=r_{\mathrm{c}}$. To analyze this critical trajectory near $r=r_{\mathrm{c}}$, we introduce a small dimensionless deviation $\delta_\ast(\varphi)$ of the areal radius by writing $R(r(\varphi))=R_{\mathrm{c}} \bigl[1+\delta_\ast(\varphi)\bigr]$, with $|\delta_\ast|\ll1$. Using $\mathrm{d}r/\mathrm{d}\varphi=(1/R')(\mathrm{d}R/\mathrm{d}\varphi)$, Eq.~\eqref{eq:orbitaleq} can be rewritten in terms of $R$ as 
\begin{align}
\left(
\frac{\mathrm{d}R}{\mathrm{d}\varphi}
\right)^2+(R')^2\:\!V(r)=0.
\label{eq:orbitaleqR}
\end{align}
Expanding $(R')^2V(r)$ about $\delta_\ast=0$ (i.e., $R=R_{\mathrm{c}}$) and using the marginal-orbit conditions in Eq.~\eqref{eq:mupco}, we obtain
\begin{align}
(R')^2\:\!V(r)= \frac{R_{\mathrm{c}}^3}{6R'_{\mathrm{c}}}V'''_{\mathrm{c}} \delta_\ast^3+O(|\delta_\ast|^4).
\label{eq:Rp2Vexp}
\end{align}
Substituting Eq.~\eqref{eq:Rp2Vexp} into Eq.~\eqref{eq:orbitaleqR} yields, to leading order,
\begin{align}
\biggl(
\frac{\mathrm{d}\delta_\ast}{\mathrm{d}\varphi}
\biggr)^2\simeq
\kappa
\:\!\delta_\ast^3,
\qquad 
\kappa \equiv -\frac{R_{\mathrm{c}}}{R'_{\mathrm{c}}} \frac{V'''_{\mathrm{c}}}{6}. 
\label{eq:localdev}
\end{align}
We find that $\delta_\ast>0$ [i.e., $R(\varphi)>R_{\mathrm{c}}$] for the critical trajectory and $\kappa>0$ since $V'''_{\mathrm{c}}<0$ as required in Eq.~\eqref{eq:mupco} and $R'_{\mathrm{c}}>0$ in our gauge. In this regime, Eq.~\eqref{eq:localdev} yields the asymptotic behavior,
\begin{align}
\delta_\ast(\varphi)\simeq \frac{4}{\kappa}(\varphi-\varphi_0)^{-2},
\label{eq:localsol}
\end{align}
where $\varphi_0$ is an integration constant. Since $\delta_\ast\to 0$ only as $|\varphi-\varphi_0|\to\infty$, the critical trajectory winds around the marginally unstable circular orbit infinitely many times as it approaches $r=r_{\mathrm{c}}$. Hence, the approach to $r=r_{\mathrm{c}}$ follows a power-law behavior rather than the exponential behavior found in the generic unstable case ($V''_{\mathrm{c}}<0$); see, e.g., Ref.~\cite{Igata:2026hzb}. The power-law exponent in Eq.~\eqref{eq:localsol} follows from the scaling symmetry of Eq.~\eqref{eq:localdev}. Under the scaling $\varphi\to \eta \varphi$ and $\delta_\ast\to \eta^{-2}\delta_\ast$, both terms scale as $\eta^{-6}$, which fixes $\delta_\ast\propto(\varphi-\varphi_0)^{-2}$.

The local solution~\eqref{eq:localsol} fixes the near-orbit asymptotics of the critical trajectory and thereby suggests a power-law (rather than logarithmic) divergence of the deflection angle in the SDL. However, as shown in the next section, this local analysis determines only the scaling exponent. The divergent coefficient (as well as the finite part) must be extracted from the deflection integral in Eq.~\eqref{eq:deflectionangle}. We therefore isolate the divergence by a near-critical expansion of the deflection integral in Sec.~\ref{sec:4}.

To characterize the local structure of the effective potential beyond the marginal conditions, we compute the third derivative $V'''_{\mathrm{c}}$ at the marginally unstable circular orbit. Using Eqs.~\eqref{eq:bcsq} and~\eqref{eq:ccond}, together with Eq.~\eqref{eq:Vppc2}, we eliminate $b$, $A'_{\mathrm{c}}$, and $A''_{\mathrm{c}}$ and obtain
\begin{align}
V'''_{\mathrm{c}}=-\frac{2R_{\mathrm{c}}^2}{B_{\mathrm{c}}}\left(
\frac{R'''_{\mathrm{c}}}{R_{\mathrm{c}}}+3\frac{R'_{\mathrm{c}}R''_{\mathrm{c}}}{R_{\mathrm{c}}^2}-\frac{A'''_{\mathrm{c}}}{2A_{\mathrm{c}}}
\right).
\label{eq:Vpppc}
\end{align}
This expression shows that the leading nonvanishing term in the expansion of $V(r)$ near the marginally unstable circular orbit is determined by the metric functions and their derivatives evaluated at $r=r_{\mathrm{c}}$. As shown in the next section, $V'''_{\mathrm{c}}$ plays a central role in determining the leading divergence of the deflection angle in the SDL.

\section{Deflection angle in the strong deflection limit}
\label{sec:4}
We now consider the SDL, defined as the limit in which the turning point approaches the marginally unstable circular photon orbit, $R_0 \to R_{\mathrm{c}}$. Equivalently, Eq.~\eqref{eq:bsq} shows that this corresponds to $b \to b_{\mathrm{c}}$, where $b_{\mathrm{c}}$ is the critical impact parameter defined in Eq.~\eqref{eq:bcsq}.

In what follows, we analyze the deflection angle in this limit. To isolate the divergent contribution, we introduce an integration variable that measures the deviation from the turning point ($R=R_0$). We define $z\equiv 1-R_0/R(r)$. With this choice, the turning point corresponds to $z=0$ and infinity to $z=1$. Since $R'(r)>0$, this transformation is one to one along a scattering trajectory and maps $r\in [r_0,\infty)$ to $z\in [0,1]$. Here and below, functions of $r$ are understood as functions of $z$ via $r=r(z;R_0)$ along the scattering trajectory.
Using $z$, the deflection integral becomes
\begin{align}
\Delta \varphi(R_0)=2 \int_0^1 f(z;R_0)\:\!\mathrm{d}z,
\end{align}
where the integrand reads
\begin{align}
f(z; R_0)=\left[\:\!
(1-z)^4\frac{(R')^2 (-V)}{R_0^2}
\:\!\right]^{-1/2}.
\end{align}

To isolate the divergence of $\Delta \varphi(R_0)$ as $R_0\to R_{\mathrm{c}}$, we expand the argument of the square root in $f(z;R_0)$ about $z=0$. 
Retaining terms through $O(z^3)$, we obtain the divergent part of the integral in the form
\begin{align}
I_{\mathrm{D}}(R_0)=2\int_0^1 \frac{\mathrm{d}z}{\sqrt{\zeta_1 z+\zeta_2 z^2+\zeta_3z^3}},
\label{eq:ID}
\end{align}
where the coefficients are given by
\begin{align}
\zeta_1&=-\frac{R'_0 V'_0}{R_0},
\label{eq:zeta1}
\\
\zeta_2&=3 \left(
\frac{R'_0}{R_0}-\frac{R''_0}{2R'_0}
\right)V'_0-\frac{V''_0}{2},
\\
\zeta_3&=\left(
-\frac{3R'_0}{R_0}+\frac{3R''_0}{R'_0}
+\frac{R_0 (R''_0)^2}{2(R'_0)^3}
-\frac{5R_0R'''_0}{6(R'_0)^2}
\right) V'_0+\left(
1-\frac{R_0 R''_0}{2(R'_0)^2}
\right)V''_0-\frac{R_0}{6R'_0}V'''_0.
\label{eq:zeta3}
\end{align}
For scattering trajectories, the turning point $r=r_0$ satisfies $V(r_0)=0$. Since photons are confined to $V\le0$ and the trajectory extends to $r>r_0$, we must have $V'_0<0$. Equality holds only in the critical limit $r_0\to r_{\mathrm{c}}$. With $R'(r)>0$, Eq.~\eqref{eq:zeta1} gives $\zeta_1>0$, and hence $\zeta_1\to0^+$ as $r_0\to r_{\mathrm{c}}$.

To quantify the proximity to the SDL, we introduce the dimensionless parameter
\begin{align}
\delta=\frac{R_0}{R_{\mathrm{c}}}-1.
\label{eq:delta}
\end{align}
We label the two near-critical branches by the sign parameter $s \equiv \operatorname{sgn}(\delta)\in\{+1,-1\}$ (i.e., $\delta = s\,|\delta|$). Here, $|\delta|$ measures the fractional deviation of $R_0$ from $R_{\mathrm{c}}$, and $s=+1$ ($s=-1$) corresponds to a radial turning point outside (inside) the marginally unstable circular orbit, $R_0>R_{\mathrm{c}}$ ($R_0<R_{\mathrm{c}}$). Depending on the structure of $V(r)$, both branches may be realized~\cite{Shaikh:2019itn,Tsukamoto:2021fsz}. For any branch-dependent quantity $X_s$, we use the shorthand $X_{+}\equiv X_{s=+1}$ and $X_{-}\equiv X_{s=-1}$.

To obtain the near-critical expansion of the coefficients $\zeta_i$, we relate $r_0-r_{\mathrm{c}}$ to $\delta$. Since $R_0=R_{\mathrm{c}}(1+\delta)$ and $R'_{\mathrm{c}}>0$, expanding $R(r_0)$ about $r_0=r_{\mathrm{c}}$ gives $r_0-r_{\mathrm{c}}=(R_{\mathrm{c}}/R'_{\mathrm{c}}) \delta+O(|\delta|^2)$. Taylor expanding $V'(r_0)$, $V''(r_0)$, and $V'''(r_0)$ about $r_0=r_{\mathrm{c}}$ and using $V'_{\mathrm{c}}=0$ and $V''_{\mathrm{c}}=0$, we obtain
\begin{align}
V'_0=\frac{R_{\mathrm{c}}^2 V'''_{\mathrm{c}}}{2(R'_{\mathrm{c}})^2} \delta^2+O(|\delta|^3),
\qquad
V''_0=\frac{R_{\mathrm{c}} V'''_{\mathrm{c}}}{R'_{\mathrm{c}}} \delta+O(|\delta|^2), 
\qquad
V'''_0=V'''_{\mathrm{c}}+O(|\delta|).
\end{align}
Substituting these into Eqs.~\eqref{eq:zeta1}--\eqref{eq:zeta3}, we obtain
\begin{align}
\zeta_1&=
3\kappa \delta^2+O(|\delta|^3),
\quad
\zeta_2=3\kappa
 \delta+O(|\delta|^2),
\quad
\zeta_3=
\kappa+O(|\delta|),
\label{eq:cexp}
\end{align}
where $\kappa$ is the coefficient introduced in Eq.~\eqref{eq:localdev} and is given by
\begin{align}
\kappa=-\frac{R_{\mathrm{c}}}{R'_{\mathrm{c}}} \frac{V'''_{\mathrm{c}}}{6}.
\label{eq:kappa}
\end{align}
For scattering trajectories, we have $V'_0<0$ and hence $\zeta_1>0$ [Eq.~\eqref{eq:zeta1}]. Comparing this with the near-critical expansion $\zeta_1=3\kappa \delta^2+O(|\delta|^3)$ implies $\kappa >0$, i.e., $V'''_{\mathrm{c}}<0$ in our radial gauge ($R'_{\mathrm{c}}>0$). As discussed in Sec.~\ref{sec:3}, $\kappa>0$ means that the critical $b=b_{\mathrm{c}}$ trajectory itself approaches the marginal orbit from a larger areal radius, so it can be reached as the limit of the scattering trajectories considered here. This condition is independent of the branch label $s$, which instead distinguishes the two near-critical turning-point branches, $R_0>R_{\mathrm{c}}$ and $R_0<R_{\mathrm{c}}$. Substituting Eq.~\eqref{eq:cexp} into Eq.~\eqref{eq:ID}, we obtain
\begin{align}
I_{\mathrm{D}}(R_0)
&=\frac{1}{\sqrt{\kappa}}\int_0^1 \frac{2+O(|\delta|)}{\sqrt{3 \delta^2z +3 \delta z^2+ z^3}}\:\!\mathrm{d}z.
\label{eq:IDexp}
\end{align}
The dominant contribution comes from the near-critical region $z=O(|\delta|)$, in which $\delta^2z$, $\delta z^2$, and $z^3$ are all $O(|\delta|^3)$. We therefore rescale $z = |\:\!\delta\:\!| \:\!y$. Writing $\delta = s|\delta|$, this yields
\begin{align}
I_{\mathrm{D}}(R_0)=\frac{1}{\sqrt{\kappa\:\! |\:\!\delta\:\!|}} \int_0^{1/|\:\!\delta\:\!|}
F_s(y)
\:\!\mathrm{d}y
+O(|\:\!\delta\:\!|^{1/2}),
\label{eq:IDF}
\end{align}
where $F_s(y)$ is defined by
\begin{align}
F_s(y)=\frac{2}{\sqrt{y\left(y^2+3\:\!s\:\!y+3\right)}}.
\label{eq:Fdef}
\end{align}
Note that $\int_0^{1/|\:\!\delta\:\!|}F_s(y)\mathrm{d}y=O(1)$ as $|\delta|\to 0$, since $F_s(y)\sim y^{-3/2}$ as $y\to \infty$. Consequently, the $O(\delta)$ corrections neglected in the integrand contribute only at $O(|\delta|^{1/2})$ in Eq.~\eqref{eq:IDF}.

To extract both the leading divergence and the constant term in the SDL, we split the integral into two parts:
\begin{align}
I_{\mathrm{D}}(R_0)=\frac{1}{\sqrt{\kappa\:\!|\:\!\delta\:\!|}}
\left(
\int_0^\infty F_s(y)\:\!
\mathrm{d}y-\int_{1/|\delta|}^\infty F_s(y)
\:\!\mathrm{d}y
\right)+O(|\:\!\delta\:\!|^{1/2}).
\end{align}
The first term gives the leading divergence, whereas the second term captures the large-$y$ tail. 
As shown below, this tail contributes a finite constant as $|\delta|\to0$. We therefore define $\mathscr{U}_s$ and $\bar{d}_{\mathrm{D}}$ by
\begin{align}
\mathscr{U}_s&=\int_0^\infty F_s(y)\:\!\mathrm{d}y,
\label{eq:Us}
\\
\bar{d}_{\mathrm{D}}&=-\lim_{|\delta|\to 0}\frac{1}{\sqrt{\kappa\:\!|\:\!\delta\:\!|}}\int_{1/|\delta|}^\infty F_s(y)\:\!\mathrm{d}y.
\end{align}
With these definitions, the divergent part of the deflection integral can be written as
\begin{align}
I_{\mathrm{D}}(R_0)=\frac{
\mathscr{U}_{s}
}{\sqrt{\kappa\:\!|\:\!\delta\:\!|}}+\bar{d}_{\mathrm{D}}+O(|\:\!\delta\:\!|^{1/2}).
\end{align}

We now evaluate $\mathscr{U}_s$ and $\bar{d}_{\mathrm{D}}$. Equation~\eqref{eq:Us} yields universal constants, explicitly given by
\begin{align}
\mathscr{U}_{+}&
=\mathscr{U},
\label{eq:Up}
\\
\mathscr{U}_{-}&=\sqrt{3}\, \mathscr{U},
\label{eq:Um}
\end{align}
where we define $\mathscr{U}$ as 
\begin{align}
\mathscr{U}\equiv \int_0^\infty F_+(y)\:\!\mathrm{d}y=\frac{2\sqrt{\pi}}{3}\frac{\Gamma(\frac{1}{6})}{\Gamma(\frac{2}{3})}.
\label{eq:mathscrU}
\end{align}
Here, $\Gamma$ denotes the gamma function. Numerically, $\mathscr{U}\approx 4.85730$ (hence, $\mathscr{U}_-=\sqrt{3}\,\mathscr{U}\approx 8.41309$). The finite term $\bar{d}_{\mathrm{D}}$ comes from the large-$y$ tail of the integral.
Using the asymptotic behavior $F_s(y)\simeq 2y^{-3/2}$ as $y\to\infty$,
we obtain
\begin{align}
\bar{d}_{\mathrm{D}}
=
-\frac{4}{\sqrt{\kappa}}.
\label{eq:dbD}
\end{align}
Because $F_s(y)$ has the same large-$y$ asymptotics for $s=\pm1$, $\bar{d}_{\mathrm{D}}$ is branch independent.

Having isolated the divergent contribution $I_{\mathrm{D}}(R_0)$, we define the remaining regular part by
\begin{align}
\Delta \varphi(R_0)=I_{\mathrm{D}}(R_0)+I_{\mathrm{R}}(R_0).
\end{align}
The regular remainder is
\begin{align}
I_{\mathrm{R}}(R_0) =2\int_0^1 \left[\:\!
f(z; R_0)-\frac{1}{\sqrt{\zeta_1 z+\zeta_2 z^2+\zeta_3z^3}}
\:\!\right]
\:\!\mathrm{d}z.
\label{eq:IR}
\end{align}
By construction, the square-bracket integrand in Eq.~\eqref{eq:IR} has a finite limit as $z\to0$ (the leading $z^{-1/2}$ divergence cancels), so $I_{\mathrm{R}}(R_0)$ is well defined. Indeed, expanding the argument of the square root shows that $f(z;R_0)=1/\sqrt{\zeta_1 z+\zeta_2 z^2+\zeta_3 z^3}+O(z^{5/2})$ as $z\to0$. We denote this limit by
\begin{align}
d_{\mathrm{R}}\equiv \lim_{R_0\to R_{\mathrm{c}}} I_{\mathrm{R}}(R_0).
\label{eq:dR}
\end{align}
For a given spacetime, $d_{\mathrm{R}}$ is obtained by evaluating Eq.~\eqref{eq:IR} in the limit $R_0\to R_{\mathrm{c}}$ (typically numerically), whereas $c_s$ is fixed by local data at $r=r_{\mathrm{c}}$.

Combining $I_{\mathrm{D}}(R_0)$ with $d_{\mathrm{R}}$, the deflection angle in the SDL takes the form
\begin{align}
\hat{\alpha}(R_0)
=c_{s}\,|\:\!\delta\:\!|^{-1/2} +d+O(|\:\!\delta\:\!|^{1/2}),
\label{eq:alphaSDL}
\end{align}
where 
\begin{align}
c_{s}
&=\frac{\mathscr{U}_{s}}{\sqrt{\kappa}},
\label{eq:Cmpm}
\\
d&=\bar{d}_{\mathrm{D}}+d_{\mathrm{R}}-\pi.
\label{eq:dfinal}
\end{align}
In the SDL ($\delta\to 0$), the deflection angle diverges as $|\delta|^{-1/2}$ when the turning point approaches the marginally unstable circular orbit. The exponent $-1/2$ is the manifestation, at the level of the integral, of the leading cubic behavior of $V(r)$ near $r=r_{\mathrm{c}}$. The divergent coefficient factorizes as $c_s=\mathscr{U}_s/\sqrt{\kappa }$. Here, $\kappa $ is determined by local geometric data at the marginally unstable circular orbit, while $\mathscr{U}_s$ are universal constants encoding the near-region integral and corresponding to the two branches $s=+1$ ($\delta>0$) and $s=-1$ ($\delta<0$). In contrast, the constant term $d$ depends on the global structure of the spacetime through the regular remainder of the deflection integral.

This structure also clarifies the relation to the local analysis of Sec.~\ref{sec:3}. Retaining only the cubic term in the expansion of $V(r)$ near $r=r_{\mathrm{c}}$ yields the scaling estimate $\hat{\alpha}_{\mathrm{loc}}\simeq (4/\sqrt{\kappa})|\delta_\ast|^{-1/2}$, where $\delta_\ast$ measures the near-orbit deviation along the critical trajectory. The local analysis therefore fixes the exponent but not the normalization. The full near-region integral replaces the naive prefactor $4$ by the branch-dependent universal constants $\mathscr{U}_s$ and also supplies the finite constant $d$, which is inaccessible to the purely local solution. Numerically, $\mathscr{U}_+\approx 4.857$ is close to the naive prefactor $4$, but for the $\delta<0$ branch, one has $\mathscr{U}_-=\sqrt{3}\,\mathscr{U}_+\simeq 8.413$, showing that the branch dependence is quantitatively essential.

To express the SDL in terms of the impact parameter $b$, we relate $\delta$ to the deviation of $b$ from its critical value. Expanding Eq.~\eqref{eq:bsq}, $b^2=R_0^2/A_0$, about $r_0=r_{\mathrm{c}}$ and using Eqs.~\eqref{eq:ccond} and~\eqref{eq:Vppc2}, we find that the linear and quadratic terms vanish. Since $r_0-r_{\mathrm{c}}=(R_{\mathrm{c}}/R'_{\mathrm{c}})\delta+O(|\delta|^2)$, this yields 
\begin{align}
\frac{b^2}{b_{\mathrm{c}}^2}=1-\frac{B_{\mathrm{c}}R_{\mathrm{c}}V'''_{\mathrm{c}}}{6(R'_{\mathrm{c}})^3} \delta^3+O(|\delta|^4).
\end{align}
Taking the square root, we obtain
\begin{align}
\frac{b}{b_{\mathrm{c}}}=
1+\mathcal{K} \:\!\delta^3+O(|\delta|^4),
\label{eq:bdelta}
\end{align}
where 
\begin{align}
\mathcal{K} \equiv 
-\frac{B_{\mathrm{c}}R_{\mathrm{c}}V'''_{\mathrm{c}} }{12 (R'_{\mathrm{c}})^3} 
=\frac{\kappa }{2} \left(
1-\frac{2m_{\mathrm{c}}}{R_{\mathrm{c}}}
\right)^{-1}>0.
\label{eq:Kappa}
\end{align}
The inequality $\mathcal{K}>0$ follows from $\kappa>0$ and $1-2m_{\mathrm{c}}/R_{\mathrm{c}}=(R'_{\mathrm{c}})^2/B_{\mathrm{c}}>0$ [Eq.~\eqref{eq:MS}].

We introduce a dimensionless parameter $\varepsilon$ measuring the fractional deviation of $b$ from $b_{\mathrm{c}}$, defined by
\begin{align}
\varepsilon=\frac{b}{b_{\mathrm{c}}}-1.
\label{eq:varepsilon}
\end{align}
Equation~\eqref{eq:bdelta} implies that $\delta \to 0$ is equivalently described by
$\varepsilon \to 0$, with the leading-order relation
\begin{align}
\varepsilon = \mathcal{K}\,\delta^3 + O(|\delta|^4).
\label{eq:epsilon-delta}
\end{align}
Inverting Eq.~\eqref{eq:epsilon-delta} gives $|\delta|=(|\varepsilon|/\mathcal{K})^{1/3}+O(|\varepsilon|^{2/3})$. Since $\mathcal{K}>0$, we also have $\operatorname{sgn}(\varepsilon)=\operatorname{sgn}(\delta)=s$. Substituting this into Eq.~\eqref{eq:alphaSDL} expresses the deflection angle in the SDL in terms of $b$ as
\begin{align}
\hat{\alpha}(b)
=\bar{c}_{s}\,|\varepsilon|^{-1/6}+d+O(|\:\!\varepsilon\:\!|^{1/6}),
\label{eq:alphaSDLb}
\end{align}
where 
\begin{align}
\bar{c}_s&=c_s \mathcal{K}^{1/6}
=\mathscr{U}_s\left[\:\!2 \:\!\kappa^{2} \left(1-\frac{2m_{\mathrm{c}}}{R_{\mathrm{c}}}\right)\:\!\right]^{-1/6}.
\label{eq:cbpm}
\end{align}
This expression shows that the $-1/6$ power-law divergence of the deflection angle is a direct consequence of the cubic behavior of the effective potential near the marginally unstable circular orbit.

\section{Geometrical characterization of marginal instability}
\label{sec:5}
In this section, we rewrite the marginal-instability conditions and the SDL coefficients in terms of local curvature in a static orthonormal tetrad and, within general relativity, in terms of the energy density and the principal pressures (radial and tangential).

We introduce an orthonormal tetrad adapted to static observers,
\begin{align}
e_{(0)}=\frac{1}{\sqrt{A(r)}}\,\partial_t,
\qquad
e_{(1)}=\frac{1}{\sqrt{B(r)}}\,\partial_r,
\qquad
e_{(2)}=\frac{1}{R(r)}\,\partial_\theta,
\qquad
e_{(3)}=\frac{1}{R(r) \sin \theta}\,\partial_\varphi.
\label{eq:staticframe}
\end{align}
In what follows, we project the curvature onto this tetrad and work with the corresponding tetrad components. The electric part of the Weyl tensor $C_{abcd}$ measured by these observers is defined as $E_{(i)(j)} = C_{abcd}e^a_{(i)}e^b_{(0)}e^c_{(j)}e^d_{(0)}$. In spherical symmetry, $E_{(i)(j)}$ takes the diagonal form
\begin{align}
E_{(i)(j)}=\mathcal{E}(r)\, \mathrm{diag}\left(
-2,\, 1,\, 1
\right).
\end{align}
Thus, the electric Weyl tensor is fully determined by the single scalar $\mathcal{E}(r)$. Similarly, the tetrad components of the Einstein tensor $G_{ab}=R_{ab}-(\mathcal{R}/2)g_{ab}$ (with scalar curvature $\mathcal{R}\equiv g^{ab}R_{ab}$) are defined as $G_{(\mu)(\nu)} = G_{ab}\, e_{(\mu)}^{a} e_{(\nu)}^{b}$. Here, parentheses denote tetrad indices; Latin indices $i,j=1,2,3$ label spatial components; and Greek indices $\mu,\nu=0,1,2,3$ run over all tetrad components.

With $m(r)$ denoting the Misner--Sharp mass defined in Eqs.~\eqref{eq:MSdef}--\eqref{eq:MS}, the scalar $\mathcal{E}(r)$ satisfies
\begin{align}
R(r)^2\:\!\mathcal{E}(r)=\frac{m(r)}{R(r)}-\frac{R(r)^2}{6}(G_{(0)(0)}+G_{(3)(3)}-G_{(1)(1)}),
\label{eq:RsqE}
\end{align}
which is a purely geometrical identity following from the decomposition of the Riemann tensor into Weyl and Ricci parts together with Eq.~\eqref{eq:MS}. For later use, we also provide the identity
\begin{align}
G_{(1)(1)}'=\frac{2R'(r)}{R(r)}\bigl(G_{(3)(3)}-G_{(1)(1)}\bigr)-\frac{A'(r)}{2A(r)}\bigl(G_{(0)(0)}+G_{(1)(1)}\bigr),
\label{eq:cBid}
\end{align}
which follows from the contracted Bianchi identity, $\nabla_a G^{ab}=0$.

Using the circular-orbit condition~\eqref{eq:ccond} together with the marginal condition~\eqref{eq:Vppc2} (see the Appendix for details), we obtain
\begin{align}
R_{\mathrm{c}}^2\:\!\mathcal{E}_{\mathrm{c}}=\frac{1}{6}.
\label{eq:RsqE16}
\end{align}
We also have
\begin{align}
R_{\mathrm{c}}^2\:\!G_{(0)(0)}^{\:\!\mathrm{c}}&=\frac{2m'_{\mathrm{c}}}{R'_{\mathrm{c}}},
\label{eq:G00c}
\\
1+R_{\mathrm{c}}^2\:\!G_{(1)(1)}^{\:\!\mathrm{c}}&=3\left(
1-\frac{2m_{\mathrm{c}}}{R_{\mathrm{c}}}\right),
\label{eq:G11ctr}
\\
R_{\mathrm{c}}^2\:\!G_{(2)(2)}^{\:\!\mathrm{c}}&=R_{\mathrm{c}}^2\:\!G_{(3)(3)}^{\:\!\mathrm{c}}=1-\frac{2m'_{\mathrm{c}}}{R'_{\mathrm{c}}}.
\label{eq:G33c}
\end{align}

We next show that the expression for $V'''_{\mathrm{c}}$ in Eq.~\eqref{eq:Vpppc} can be rewritten in terms of local curvature quantities. Specifically, we find
\begin{align}
V'''_{\mathrm{c}}=-12 \left[\:\!
R^2(r)\:\! \mathcal{E}(r)
\:\!\right]'_{\mathrm{c}}.
\label{eq:VpppcE}
\end{align}
We can interpret $R(r)^2\mathcal{E}(r)$ as a dimensionless measure of the tidal field encoded in the electric part of the Weyl tensor. With this identification, Eq.~\eqref{eq:VpppcE} shows that $V'''_{\mathrm{c}}$, which governs the leading cubic behavior of the effective potential near a marginally unstable circular photon orbit, is governed by the coordinate-radius derivative of this local tidal field (in the chosen radial gauge), evaluated at $r=r_{\mathrm{c}}$. Equivalently, combining Eqs.~\eqref{eq:RsqE} and~\eqref{eq:cBid}, we can rewrite Eq.~\eqref{eq:VpppcE} in terms of the Einstein tensor components as
\begin{align}
V'''_{\mathrm{c}}=2\mathcal{N}'_{\mathrm{c}},
\end{align}
where $\mathcal{N}(r)$ is defined as%
\footnote{In the nonmarginal case, the logarithmic coefficient of the deflection angle in the SDL behaves as $\bar{a}=1/\sqrt{1-\mathcal{N}_{\mathrm{c}}} $~\cite{Igata:2025taz}; it becomes singular in the marginal limit since Eqs.~\eqref{eq:G00c} and~\eqref{eq:G33c} imply $\mathcal{N}_{\mathrm{c}}=1$.}
\begin{align}
\mathcal{N}(r)\equiv R^2(r) \bigl(
G_{(0)(0)}+G_{(3)(3)}
\bigr).
\label{eq:N}
\end{align}
Substituting this expression into Eqs.~\eqref{eq:kappa} and~\eqref{eq:cbpm} yields
\begin{align}
c_s&=\frac{\mathscr{U}_{s}}{\sqrt{\kappa}},
\label{eq:c3SDL2}
\\
\bar{c}_s&
=
\mathscr{U}_{s} \left[\:\!
\frac{2}{3}\kappa^2\big(1+R_{\mathrm{c}}^2 G_{(1)(1)}^{\:\!\mathrm{c}}\big)
\:\!\right]^{-1/6},
\label{eq:cbpm2}
\end{align}
where we have used Eq.~\eqref{eq:G11ctr}, and $\kappa$ is given by
\begin{align}
\kappa &=\frac{R_{\mathrm{c}}}{3} 
\Bigl(-\frac{\mathrm{d}\mathcal{N}}{\mathrm{d}R}\Bigr)_{\!\mathrm{c}}.
\label{eq:kappaN}
\end{align}
In deriving Eq.~\eqref{eq:kappaN}, we rewrite derivatives with respect to $r$ in terms of those with respect to the areal radius $R$, i.e., $\mathrm{d}/\mathrm{d}R=(1/R'(r))\,\mathrm{d}/\mathrm{d}r$. This is well defined because we have fixed the radial gauge such that $R'(r)>0$. Apart from the branch-dependent universal factor $\mathscr{U}_s$, Eq.~\eqref{eq:c3SDL2} shows that $c_s$ is fixed locally by $\kappa$, or equivalently by the slope $(\mathrm{d}\mathcal{N}/\mathrm{d}R)_{\mathrm{c}}$ of the weighted curvature combination $\mathcal{N}$ together with the areal radius $R_{\mathrm{c}}$. Equation~\eqref{eq:cbpm2} further shows that $\bar{c}_s$ depends on the same local data and, in addition, on the combination $1+R_{\mathrm{c}}^2 G_{(1)(1)}^{\:\!\mathrm{c}}$, which enters through the conversion from $\delta$ to $\varepsilon$.

\subsection{Stress-energy characterization of marginal instability}
We specialize the geometrical relations of Sec.~\ref{sec:5} to general relativity. In the orthonormal frame~\eqref{eq:staticframe} of static observers, the Einstein equations read $G_{(\mu)(\nu)} = 8\pi T_{(\mu)(\nu)}$, where $T_{(\mu)(\nu)}$ is the stress-energy tensor. In this frame, the stress-energy tensor is diagonal, 
\begin{align}
T_{(\mu)(\nu)}=\mathrm{diag}\left[\:\!
\rho(r),\,P(r),\,\Pi(r),\,\Pi(r)
\:\!\right],
\end{align}
where $\rho$ is the energy density, $P$ is the radial pressure, and $\Pi$ is the tangential pressure. In terms of these quantities, Eq.~\eqref{eq:N} becomes
\begin{align}
\mathcal{N}(r)= 8\pi R(r)^2\bigl[\:\!
\rho(r)+\Pi(r)
\:\!\bigr].
\label{eq:Nmat}
\end{align}

To interpret $\rho+\Pi$, we introduce a null vector tangent to the circular photon orbits, $k=e_{(0)}+e_{(3)}$ [i.e., $k^{(\mu)}=(1,0,0,1)$]. Then, the contraction $T_{(\mu)(\nu)}k^{(\mu)}k^{(\nu)}$ yields 
\begin{align}
T_{(\mu)(\nu)}k^{(\mu)}k^{(\nu)}=\rho+\Pi. 
\end{align}
This means that $\rho+\Pi$ is the null-energy density along $k$. Using the Einstein equations and the null condition $k_{(\mu)}k^{(\mu)}=0$, this is also the Ricci focusing term in the Raychaudhuri equation, $R_{(\mu)(\nu)}k^{(\mu)}k^{(\nu)}=8\pi (\rho+\Pi)$.

At the marginally unstable circular orbit, Eqs.~\eqref{eq:G00c} and~\eqref{eq:G33c} imply $\mathcal{N}_{\mathrm{c}}=1$.
In general relativity, this translates into the local condition
\begin{align}
8\pi R_{\mathrm{c}}^2\bigl(\rho_{\mathrm{c}}+\Pi_{\mathrm{c}}\bigr)=1,
\label{eq:marginalcondmatter}
\end{align}
which fixes the value of the null-energy density along the photon-orbit direction at $r=r_{\mathrm{c}}$~\cite{Hod:2017zpi}.
The remaining information relevant for the SDL is encoded in the slope of Eq.~\eqref{eq:Nmat} at the marginally unstable circular orbit. Indeed, combining Eqs.~\eqref{eq:c3SDL2},~\eqref{eq:kappaN}, and~\eqref{eq:Nmat}, we can write
\begin{align}
c_{s}=\frac{\mathscr{U}_{s}}{\sqrt{\kappa}},
\qquad
\kappa
= -\frac{8\pi R_{\mathrm{c}}}{3} \left(
\frac{\mathrm{d}}{\mathrm{d}R}
\left[\:\!
R^2\left(\rho+\Pi\right)\:\!
\right]\right)_{\!\mathrm{c}}.
\label{eq:kappamatter}
\end{align}
The condition $\kappa>0$ required for near-critical scattering trajectories [see the discussion below Eq.~\eqref{eq:kappa}] implies 
\begin{align}
\left(
\frac{\mathrm{d}}{\mathrm{d}R}
\left[\:\!
R^2\left(\rho+\Pi\right)\:\!
\right]\right)_{\!\mathrm{c}}<0.
\end{align}
Therefore, the coefficient of the power-law divergence of $\hat{\alpha}(R_0)$ in Eq.~\eqref{eq:alphaSDL} is controlled by how rapidly the dimensionless null-energy density $R^2(\rho+\Pi)$ decreases with $R$ near the marginally unstable circular photon orbit.

Using the Einstein equation
$G_{(1)(1)}=8\pi P$, Eq.~\eqref{eq:cbpm2} becomes
\begin{align}
\bar{c}_s&
=
\mathscr{U}_{s} \left[\:\!
\frac{2}{3}\kappa^2\bigl(1+8\pi R_{\mathrm{c}}^2 P_{\:\!\mathrm{c}}\bigr)
\:\!\right]^{-1/6}.
\label{eq:cbar_matter}
\end{align}
This shows explicitly that (i) the dimensionless null-energy density $R^2(\rho+\Pi)$ affects $\bar{c}_s$ only through the local slope of the radial profile encoded in $\kappa$ and (ii) $P_{\mathrm{c}}$ controls the $\delta$--$\varepsilon$ conversion in Eq.~\eqref{eq:epsilon-delta} via the local compactness $1-2m_{\mathrm{c}}/R_{\mathrm{c}}$ at the marginally unstable circular orbit [see Eq.~\eqref{eq:G11ctr}]. In particular, at fixed $\kappa$, a larger $P_{\mathrm{c}}$ decreases $\bar{c}_s$ and thus weakens the divergence of $\hat{\alpha}(b)$ in Eq.~\eqref{eq:alphaSDLb}. Finally, note that in vacuum ($\rho=\Pi=P=0$) we have $\mathcal{N}=0$, so the marginal condition $\mathcal{N}_{\:\!\mathrm{c}}=1$ cannot be satisfied. Thus, degenerate photon spheres in general relativity necessarily require nontrivial stress energy in the neighborhood of $r=r_{\mathrm{c}}$.

\section{Applications}
\label{sec:6}
In this section, we apply the general framework developed above to specific spacetime geometries. Our aims are twofold. First, we demonstrate how the universal SDL structure for a marginally unstable circular photon orbit is realized in explicit examples. Second, we verify analytically the local relations underlying the formalism, together with the resulting leading divergent coefficients and the local near-orbit contribution $\bar{d}_{\mathrm{D}}$ to the finite term. For each example, we tune the spacetime parameters so that a photon sphere and an anti-photon sphere coalesce into a degenerate configuration. The resulting local data ($r_{\mathrm{c}}$, $b_{\mathrm{c}}$, $m_{\mathrm{c}}$, $V'''_{\mathrm{c}}$) are summarized in Table~\ref{table:1}, while the SDL inputs $(\kappa, \mathcal{K})$ and the outer-branch coefficients $(c_+, \bar{c}_+, \bar{d}_{\mathrm{D}})$ are collected in Table~\ref{table:2}. All explicit examples in this section are written in the areal-radius gauge $R(r)=r$. Throughout this section, we list the outer branch ($s=+1$); the inner branch ($s=-1$) follows universally from $c_{-}=\sqrt{3}c_+$ and $\bar{c}_{-}=\sqrt{3}\bar{c}_{+}$, and $\bar{d}_{\mathrm{D}}$ is branch independent.

\subsection{Reissner--Nordstr\"om spacetime}
We consider the Reissner--Nordstr\"om (RN) spacetime, a static and spherically symmetric solution to the Einstein--Maxwell equations. Focusing on a specific value of the electric charge, we evaluate the SDL coefficients for the marginally unstable circular photon orbit.

For the RN spacetime, the metric functions are given by
\begin{align}
A(r)=B(r)^{-1}=1-\frac{2M}{r}+\frac{Q^2}{r^2}, \quad R(r)=r, 
\end{align}
where $M$ and $Q$ denote the mass and electric charge parameters, respectively. For $Q^2/M^2\le 1$, the spacetime possesses an event horizon (extremal at equality) and a photon sphere outside it. In the subextremal case, there is no anti-photon sphere in the exterior region, while in the extremal case, it emerges on the horizon. For $1<Q^2/M^2<9/8$, the spacetime is horizonless (i.e., a naked singularity) and admits both a photon sphere and an anti-photon sphere. The two coalesce into a degenerate configuration at
\begin{align}
\frac{Q^2}{M^2}=\frac{9}{8}.
\end{align}

The values of $\bar{c}_s$ in Table~\ref{table:2} agree with the results of Refs.~\cite{Sasaki:2025web,Tsukamoto:2025hbz}, which were derived using independent approaches.

\subsection{Hayward spacetime}
We next consider the Hayward spacetime~\cite{Hayward:2005gi}, a static and spherically symmetric regular spacetime often discussed as a model of nonsingular compact objects. As in the previous example, we focus on a parameter choice for which a photon sphere and an anti-photon sphere coalesce into a degenerate photon sphere. We then evaluate the corresponding SDL coefficients.

For the Hayward spacetime, the metric functions take the form
\begin{align}
A(r)=B(r)^{-1}=1-\frac{2Mr^2}{r^3+2q^2 M}, \quad R(r)=r, 
\end{align}
where $M$ denotes the mass parameter and $q$ characterizes the deviation from the Schwarzschild geometry. For $|q|/M\le 4/(3\sqrt{3})$, the spacetime possesses an event horizon (extremal at equality) and a photon sphere outside it. In the subextremal case, there is no anti-photon sphere in the exterior region, while in the extremal case, it emerges on the horizon. For $4/(3\sqrt{3})<|q|/M<(25/24)\sqrt{5/6}$, the spacetime is horizonless and admits both a photon sphere and an anti-photon sphere. The two coalesce into a degenerate configuration at
\begin{align}
\frac{\left|\:\!q\:\!\right|}{M}
=\frac{25}{24}\sqrt{\frac{5}{6}}\approx 0.950907.
\end{align}

The value of $\bar{c}_+$ reported in Table~\ref{table:2} agrees exactly with the value implied by Ref.~\cite{Chiba:2017nml}; it is reproduced by combining their Eqs.~(15), (21), and (22). Evaluating our analytic expressions numerically, we find agreement with the numerical results in Eqs.~(3.41) and (3.43) of Ref.~\cite{Tsukamoto:2025hbz} at the $10^{-4}$ and $10^{-3}$ levels for $\bar{c}_+$ and for $\bar{c}_-=\sqrt{3}\,\bar{c}_+$, respectively.

\subsection{Bardeen spacetime}
We consider the Bardeen spacetime~\cite{Bardeen:1968,Ayon-Beato:2000mjt}, a static, spherically symmetric regular geometry often discussed as a model of nonsingular compact objects. The deflection angle in this spacetime has been previously investigated in the SDL associated with a (nondegenerate) photon sphere, where it exhibits a logarithmic divergence~\cite{Eiroa:2010wm}. Here, we focus instead on the critical configuration in which a photon sphere and an anti-photon sphere coalesce into a degenerate photon sphere, and we study the resulting power-law divergence.

For the Bardeen spacetime, the metric functions take the form
\begin{align}
A(r)=B(r)^{-1}=1-\frac{2Mr^2}{(r^2+g^2)^{3/2}}, \quad R(r)=r, 
\end{align}
where $M$ denotes the mass parameter and $g$ is a constant characterizing the deviation from the Schwarzschild geometry. For $|g|/M\le 4/(3\sqrt{3})$, the spacetime possesses an event horizon (extremal at equality) and a photon sphere outside it. In the subextremal case, there is no anti-photon sphere in the exterior region, while in the extremal case, it emerges on the horizon. For $4/(3\sqrt{3})<|g|/M<48/(25\sqrt{5})$, the spacetime is horizonless and admits both a photon sphere and an anti-photon sphere. The two coalesce into a degenerate configuration at
\begin{align}
\frac{\left|\:\!g\:\!\right|}{M}=\frac{48}{25\sqrt{5}}\approx 0.85865.
\end{align}

\subsection{Reissner--Nordstr\"om black-hole-like wormhole}
We next consider the RN black-hole-like wormhole spacetime proposed in Ref.~\cite{Lemos:2008cv}. This spacetime provides a useful example in which a degenerate photon sphere can occur outside the wormhole throat~\cite{Tsukamoto:2024pid}.

The metric functions are given by
\begin{align}
A(r) =1- \frac{2M}{(1+\lambda^2)r} + \frac{Q^2}{(1+\lambda^2)r^2},
\quad
B(r)=\left(1-\frac{2M}{r}+\frac{Q^2}{r^2}\right)^{-1},
\quad 
R(r)=r,
\end{align}
where $M>0$ is the mass parameter, $Q$ is the electric charge, and $\lambda>0$ controls the deviation from the standard RN metric. It is convenient to introduce the dimensionless charge parameter $q\equiv |Q|/M$ with $0\le q\le1$. The radius of the wormhole throat is $r_{\mathrm{th}}=M(1+\sqrt{1-q^2})$.

Requiring the existence of a degenerate photon sphere fixes $\lambda$ as
\begin{align}
\lambda=\frac{\sqrt{9-8q^2}}{2\sqrt{2}q}.
\label{eq:lambdaRNlw}
\end{align}
For this choice of parameters, the degenerate photon sphere is located at $r_{\mathrm{c}}/M=4q^2/3$. Requiring that this radius lies outside the throat, $r_{\mathrm{c}}>r_{\mathrm{th}}$, implies%
\footnote{A photon sphere is located at the throat, whereas a photon sphere and an anti-photon sphere merge on each side of the throat, forming two degenerate photon spheres in total~\cite{Tsukamoto:2024pid}.}
\begin{align}
\frac{\sqrt{15}}{4} <q\le 1.
\label{eq:qrange}
\end{align}
Equivalently, 
\begin{align}
\frac{1}{2\sqrt{2}}\le \lambda <\frac{1}{\sqrt{5}}. 
\label{eq:lambdarange}
\end{align}

\begin{table*}[t]
\caption{Degenerate photon-sphere data for the analytic examples in Sec.~\ref{sec:6}.
All quantities are evaluated at $r=r_{\mathrm{c}}$. Numbers in parentheses denote decimal approximations. For the RN-like wormhole, we use $q= |Q|/M$ and fix $\lambda$ by Eq.~\eqref{eq:lambdaRNlw}; the allowed parameter range [Eqs.~\eqref{eq:qrange} and~\eqref{eq:lambdarange}] ensures that $r_{\mathrm{c}}>r_{\mathrm{th}}$.}
\begin{ruledtabular}
\begin{tabular}{lccccc}
Spacetime & Tuning & $r_{\mathrm{c}}/M$ & $b_{\mathrm{c}}/M$ & $m_{\mathrm{c}}/M$ & $V^{\prime\prime\prime}_{\mathrm{c}} M$ \\
\hline
RN &
$Q^2/M^2=9/8$ &
$3/2$ &
$3\sqrt6/2$ $(3.67423)$ &
$5/8$ &
$-8/3$ \\
Hayward &
$\left|\:\!q\:\!\right|/M=\frac{25}{24}\sqrt{\frac{5}{6}}$ &
$25/12$ $(2.08333)$ &
$25\sqrt5/12$ $(4.65847)$ &
$5/6$ &
$-12/5$ \\
Bardeen &
$\left|\:\!g\:\!\right|/M=\frac{48}{25\sqrt5}$ &
$\frac{96}{25\sqrt5}$ $(1.71730)$ &
$\frac{96}{25}\sqrt{\frac{6}{5}}$ $(4.20651)$ &
$\frac{8}{5\sqrt5}$ &
$-\frac{5\sqrt5}{6}$ \\
RN-like wormhole &
$\lambda=\frac{\sqrt{9-8q^2}}{2\sqrt2\,q}$ &
$\frac{4}{3}q^2$ &
$\frac{4\sqrt6}{3}q^2$ &
$5/8$ &
$-\frac{9(16q^2-15)}{8q^4}$ \\
\end{tabular}
\end{ruledtabular}
\label{table:1}
\end{table*}

\begin{table*}[t]
\caption{Outer-branch SDL inputs and coefficients for the examples of Sec.~\ref{sec:6}. Numbers in parentheses denote decimal approximations.
Here, $\mathscr{U}= \frac{2\sqrt{\pi}}{3}\frac{\Gamma(1/6)}{\Gamma(2/3)}$, and the inner branch follows from $c_-=\sqrt3\,c_+$ and $\bar{c}_-=\sqrt3\,\bar{c}_+$.}
\begin{ruledtabular}
\begin{tabular}{lccccc}
Spacetime & $\kappa$ & $\mathcal{K}$ & $c_+$ & $\bar{c}_+$ & $\bar{d}_{\mathrm{D}}$ \\
\hline
RN &
$2/3$ & $2$ &
$\sqrt{\frac{3}{2}}\mathscr{U}$ $(5.94895)$ &
$\sqrt{\frac{3}{2}}\,2^{1/6}\mathscr{U}$ $(6.67748)$ &
$-2\sqrt6$ $(-4.89898)$ \\
Hayward &
$5/6$ & $25/12$ &
$\sqrt{\frac{6}{5}}\mathscr{U}$ $(5.32091)$ &
$\sqrt{\frac{6}{5}}\left(\frac{25}{12}\right)^{1/6}\mathscr{U}$ $(6.01329)$ &
$-4\sqrt{\frac{6}{5}}$ $(-4.38178)$ \\
Bardeen &
$8/15$ & $8/5$ &
$\sqrt{\frac{15}{8}}\mathscr{U}$ $(6.65113)$ &
$\sqrt{\frac{15}{8}}\left(\frac{8}{5}\right)^{1/6}\mathscr{U}$ $(7.19309)$ &
$-\sqrt{30}$ $(-5.47723)$ \\
RN-like wormhole &
$4-\frac{15}{4q^2}$ & $2$ &
$\frac{2\mathscr{U}q}{\sqrt{16q^2-15}}$ &
$\frac{2^{7/6}\mathscr{U}q}{\sqrt{16q^2-15}}$ &
$-\frac{8q}{\sqrt{16q^2-15}}$ \\
\end{tabular}
\end{ruledtabular}
\label{table:2}
\end{table*}

\section{Conclusions}
\label{sec:7}
We have developed a strong-deflection expansion for null-geodesic scattering near a degenerate photon sphere in general static and spherically symmetric spacetimes. By identifying the near-critical region that controls the divergence, rescaling the integration variable appropriately, and isolating the near-orbit contribution in the deflection-angle integral in a manner that remains nonsingular at marginality, we have obtained the leading power-law coefficient together with a decomposition of the finite offset into a universal near-region contribution and a spacetime-dependent regular remainder. Concretely, the deflection angle behaves as $\hat{\alpha}=c_s |\delta|^{-1/2} + d + O(|\delta|^{1/2})$ with $\delta=R_0/R_{\mathrm{c}} - 1$, or equivalently $\hat{\alpha}=\bar{c}_{s}\,|\varepsilon|^{-1/6}+d+O(|\:\!\varepsilon\:\!|^{1/6})$ with $\varepsilon=b/b_{\mathrm{c}}-1$. The divergent coefficient factorizes as $c_{s}=\mathscr{U}_{s}/\sqrt{\kappa}$, where the universal branch constants $\mathscr{U}_{\pm}$ correspond to the two near-critical turning-point branches, $R_0>R_{\mathrm{c}}$ and $R_0<R_{\mathrm{c}}$ ($\mathscr{U}_{-}/\mathscr{U}_{+}=\sqrt{3}$), and $\kappa$ is a single local parameter set by the first nonvanishing derivative of the effective potential evaluated at the marginally unstable circular photon orbit beyond the (vanishing) quadratic term. The offset $d$ separates into a local near-region contribution and a regular remainder sensitive to the global geometry.

We have further reformulated the relevant local input in a manifestly coordinate-invariant way using locally measurable curvature in an orthonormal tetrad. Specifically, the cubic coefficient governing the marginal instability is given by the coordinate-radius derivative of the dimensionless tidal measure $R^2\mathcal{E}$ built from the electric part of the Weyl tensor, $V'''_{\mathrm{c}}=-12(R^2 \mathcal{E})'_{\mathrm{c}}$, so the divergence is controlled by the tidal-field derivative across the marginally unstable circular photon orbit. Within general relativity, Eq.~\eqref{eq:marginalcondmatter}, first derived in Ref.~\cite{Hod:2017zpi}, already provides the local marginality condition, fixing the null-energy density along the photon-orbit direction as $8\pi R_{\mathrm{c}}^2\bigl(\rho_{\mathrm{c}}+\Pi_{\mathrm{c}}\bigr)=1$. Beyond this known condition, the present work shows that the divergent coefficient is determined by the areal-radius derivative $\kappa
= -(8\pi R_{\mathrm{c}}/3) \left(
\mathrm{d}\bigl[
R^2(\rho+\Pi)
\bigr]/\mathrm{d}R \right)_{\mathrm{c}}$ or, equivalently, by the areal-radius derivative of the tidal measure $R^2 \mathcal{E}$. The sign of $\kappa$ also determines whether near-critical scattering from infinity exists within our scattering setup. In this sense, marginal strong lensing can, in principle, probe not only the geometry at the orbit but also how the local tidal field [or the effective matter content through $R^2(\rho+\Pi)$] varies across it.

A notable feature of our extraction is the clean separation between universal and spacetime-dependent information. The two universal constants $\mathscr{U}_{\pm}$ encode the two near-critical turning-point branches, $R_0>R_{\mathrm{c}}$ and $R_0<R_{\mathrm{c}}$, and their fixed ratio $\mathscr{U}_{-}/\mathscr{U}_{+}=\sqrt{3}$ shows that the two branches are quantitatively inequivalent. This branch structure should therefore be kept explicit when relating near-critical trajectories to strong-lensing observables in marginal configurations.

Since both the SDL and eikonal quasinormal modes (QNMs) are tied to the same unstable circular photon orbit (see, e.g., reviews~\cite{Berti:2009kk,Konoplya:2011qq}), our local invariants provide concrete candidates for entering a generalized QNM-lensing correspondence in the degenerate regime. In the standard eikonal picture, the damping rate is controlled by the Lyapunov exponent~\cite{Ferrari:1984zz,Cardoso:2008bp}, which vanishes at marginality~\cite{Yang:2012pj,Yang:2012he}; because the standard formula relies on a locally quadratic barrier top in the usual Wentzel--Kramers--Brillouin (WKB) expansion about a nondegenerate maximum~\cite{Iyer:1986np}, the marginal case likely requires a uniform or modified WKB treatment around a higher-order maximum (see, e.g., Ref.~\cite{Hatsuda:2021gtn} for a review). It would be interesting to determine, via a dedicated perturbation analysis, whether $\kappa$ controls the leading damping scale and how it relates to the SDL power law. Although such a correspondence may fail in spacetimes with multiple photon spheres~\cite{Konoplya:2017wot}, spacetimes with an event horizon and an exterior degenerate photon sphere~\cite{Guo:2021enm,Guo:2023mda} may provide a clean arena for testing these ideas.

Finally, it would be valuable to extend the canonical extraction and invariant characterization beyond the static and spherically symmetric setting. In stationary and axisymmetric spacetimes, the critical structures involve families of spherical photon orbits~\cite{Teo:2003ltt} and coupled radial-polar dynamics; determining whether the marginal limit can again be treated canonically and whether the leading power-law coefficient remains local and coordinate invariant are natural next steps. More broadly, the same strategy should extend to other metric theories of gravity, with theory-specific input encoded in the field equations and the associated local curvature invariants. We leave these extensions, together with a systematic exploration of higher-order degeneracies and the associated power-law exponents, for future work.

\begin{acknowledgments}
The authors gratefully acknowledge useful comments from Shinya Aoki, Tomohiro Harada, Akihito Katsumata, Hiromi Saida, and Yohsuke Takamori. This collaboration was initiated through discussions at the Geodesics: Orbits and Rays 2025 (GOR2025) workshop. We thank the participants for their valuable comments and suggestions. T.I. was supported in part by Gakushuin University, by JSPS KAKENHI Grant Nos.~JP22K03611, JP23KK0048, JP24H00183, and JP26K07089, and by the Inamori Foundation through the Inamori Incubate Research Grants. 
\end{acknowledgments}


\appendix*

\section{Derivation of key identities for the marginally unstable circular photon orbit}
Here, we derive Eqs.~\eqref{eq:RsqE16} and~\eqref{eq:VpppcE}. We define $H(r)$ as
\begin{align}
H(r)=\log \frac{A(r)}{R(r)^2}. 
\label{eq:H}
\end{align}
The first and second derivatives of $H(r)$ are given by
\begin{align}
H'(r)&=\frac{A'(r)}{A(r)}-\frac{2R'(r)}{R(r)},
\\
H''(r)&=\frac{A''(r)}{A(r)}-\left(\frac{A'(r)}{A(r)}\right)^2-2\left[\:\!
\frac{R''(r)}{R(r)}-\left(
\frac{R'(r)}{R(r)}
\right)^2
\:\!\right]. 
\end{align}
Using Eq.~\eqref{eq:H}, the effective potential $V(r)$ in Eq.~\eqref{eq:V} can be rewritten as 
\begin{align}
V(r)=\frac{R^2(r)}{B(r)}\biggl(1-\frac{e^{-H(r)}}{b^2}\biggr).
\end{align}

We now impose the conditions for a marginally unstable circular photon orbit given in Eq.~\eqref{eq:mupco}. In terms of $H_{\mathrm{c}}\equiv H(r_{\mathrm{c}})$, $H'_{\mathrm{c}}\equiv H'(r_{\mathrm{c}})$, and $H''_{\mathrm{c}}\equiv H''(r_{\mathrm{c}})$, we have 
\begin{align}
V_{\mathrm{c}}&=\frac{R_{\mathrm{c}}^2}{B_{\mathrm{c}}}\left(1-\frac{e^{-H_{\mathrm{c}}}}{b_{\mathrm{c}}^2}\right)=0,
\label{eq:VcH}
\\
V'_{\mathrm{c}}
&=\frac{R_{\mathrm{c}}^2}{B_{\mathrm{c}}}\,H'_{\mathrm{c}}=0,
\\
V''_{\mathrm{c}}
&=\frac{R_{\mathrm{c}}^2}{B_{\mathrm{c}}}\,H''_{\mathrm{c}}=0,
\label{eq:VppcH}
\end{align}
which are equivalent to Eqs.~\eqref{eq:bcsq},~\eqref{eq:ccond}, and~\eqref{eq:Vppc2}. The quantities $H'_{\mathrm{c}}$ and $H''_{\mathrm{c}}$ are given explicitly by
\begin{align}
H'_{\mathrm{c}}
&=\frac{A'_{\mathrm{c}}}{A_{\mathrm{c}}}-\frac{2R'_{\mathrm{c}}}{R_{\mathrm{c}}},
\\
H''_{\mathrm{c}}
&=\frac{A''_{\mathrm{c}}}{A_{\mathrm{c}}}-2\left(
\frac{R'_{\mathrm{c}}}{R_{\mathrm{c}}}
\right)^2-2\frac{R''_{\mathrm{c}}}{R_{\mathrm{c}}}. 
\end{align}

Since $R_{\mathrm{c}}^2/B_{\mathrm{c}}>0$, Eqs.~\eqref{eq:VcH}--\eqref{eq:VppcH} imply
\begin{align}
b_{\mathrm{c}}^2=e^{-H_{\mathrm{c}}},
\qquad
H'_{\mathrm{c}}=0, \qquad 
H''_{\mathrm{c}}=0. 
\label{eq:hcs}
\end{align}
Using Eq.~\eqref{eq:hcs}, we obtain
\begin{align}
V'''_{\mathrm{c}}=\frac{R_{\mathrm{c}}^2}{B_{\mathrm{c}}}\,H'''_{\mathrm{c}},
\label{eq:V3_H3}
\end{align}
where
\begin{align}
H'''_{\mathrm{c}}=-2\left(
\frac{R'''_{\mathrm{c}}}{R_{\mathrm{c}}}+3\frac{R'_{\mathrm{c}}R''_{\mathrm{c}}}{R_{\mathrm{c}}^2}-\frac{A'''_{\mathrm{c}}}{2A_{\mathrm{c}}}
\right).
\end{align}

The electric part of the Weyl tensor $\mathcal{E}(r)$ can be written explicitly as 
\begin{align}
R(r)^2\:\!\mathcal{E}(r)=\frac{1}{6}-\frac{R(r)^2}{12 B(r)}\left[\:\!
H''(r)+\frac{[H'(r)]^2}{2}+H'(r) \left(
\frac{R'(r)}{R(r)}-\frac{B'(r)}{2B(r)}
\right)
\:\!\right]. 
\label{eq:REsqmet}
\end{align}
Evaluating Eq.~\eqref{eq:REsqmet} at $r=r_{\mathrm{c}}$ and using Eq.~\eqref{eq:hcs} yields $R_{\mathrm{c}}^2\:\!\mathcal{E}_{\mathrm{c}}=1/6$, in agreement with Eq.~\eqref{eq:RsqE16}. Differentiating Eq.~\eqref{eq:REsqmet} with respect to $r$ and imposing the conditions~\eqref{eq:hcs} at $r=r_{\mathrm{c}}$, we find 
\begin{align}
\left[\:\!
R(r)^2\:\!\mathcal{E}(r)
\:\!\right]'_{\mathrm{c}}=-\frac{R_{\mathrm{c}}^2}{12 B_{\mathrm{c}}}H'''_{\mathrm{c}}.
\label{eq:ddrRsqE}
\end{align}
Substituting Eq.~\eqref{eq:V3_H3} into Eq.~\eqref{eq:ddrRsqE}, we obtain 
\begin{align}
V'''_{\mathrm{c}}=-12 \,
\left[\:\!
R^2(r)\:\! \mathcal{E}(r)
\:\!\right]'_{\mathrm{c}},
\end{align}
which coincides with Eq.~\eqref{eq:VpppcE}.


\begin{thebibliography}{99}
%
\bibitem{Bardeen:1973}
J.~M.~Bardeen,
Timelike and null geodesics in the Kerr metric,
in \textit{Black Holes (Les Astres Occlus)}, edited by C.~DeWitt and B.~S.~DeWitt
(Gordon and Breach, New York, 1973), pp.~215--239.

%
\bibitem{EventHorizonTelescope:2019dse}
K.~Akiyama \textit{et al.} (Event Horizon Telescope Collaboration),
Astrophys. J. Lett. \href{https://doi.org/10.3847/2041-8213/ab0ec7}{\textbf{875}, L1 (2019)}
[arXiv:\href{http://arxiv.org/abs/1906.11238}{1906.11238 [astro-ph.GA]}].

%
\bibitem{EventHorizonTelescope:2022wkp}
K.~Akiyama \textit{et al.} (Event Horizon Telescope Collaboration),
Astrophys. J. Lett. \href{https://doi.org/10.3847/2041-8213/ac6674}{\textbf{930}, L12 (2022)}
[arXiv:\href{http://arxiv.org/abs/2311.08680}{2311.08680 [astro-ph.HE]}].

%
\bibitem{Luminet:1979}
J.~P.~Luminet,
Astron. Astrophys. \href{https://ui.adsabs.harvard.edu/abs/1979A&A....75..228L/abstract}{\textbf{75}, 228 (1979)}.

\bibitem{Fukue:1988}
J.~Fukue and T.~Yokoyama,
Publ. Astron. Soc. Jpn. \href{https://ui.adsabs.harvard.edu/abs/1988PASJ...40...15F/abstract}{\textbf{40}, 15 (1988)}.

%
\bibitem{Falcke:1999pj}
H.~Falcke, F.~Melia, and E.~Agol,
Astrophys. J. Lett. \href{https://doi.org/10.1086/312423}{\textbf{528}, L13 (2000)}
[arXiv:\href{https://arxiv.org/abs/astro-ph/9912263}{astro-ph/9912263 [astro-ph]}].

%
\bibitem{Takahashi:2004xh}
R.~Takahashi,
J. Korean Phys. Soc. \textbf{45}, S1808 (2004);
Astrophys. J. \href{https://doi.org/10.1086/422403}{\textbf{611}, 996 (2004)}
[arXiv:\href{http://arxiv.org/abs/astro-ph/0405099}{astro-ph/0405099}].

%
\bibitem{Hioki:2009na}
K.~Hioki and K.~i.~Maeda,
Phys. Rev. D \href{https://doi.org/10.1103/PhysRevD.80.024042}{\textbf{80}, 024042 (2009)}
[arXiv:\href{https://arxiv.org/abs/0904.3575}{0904.3575 [astro-ph.HE]}].

%
\bibitem{Igata:2019pgb}
T.~Igata, H.~Ishihara, and Y.~Yasunishi,
Phys. Rev. D \href{https://doi.org/10.1103/PhysRevD.100.044058}{\textbf{100}, 044058 (2019)}
[arXiv:\href{https://arxiv.org/abs/1904.00271}{1904.00271 [gr-qc]}].


%
\bibitem{Gralla:2019xty}
S.~E.~Gralla, D.~E.~Holz, and R.~M.~Wald,
Phys. Rev. D \href{https://doi.org/10.1103/PhysRevD.100.024018}{\textbf{100}, 024018 (2019)}
[arXiv:\href{https://arxiv.org/abs/1906.00873}{1906.00873 [astro-ph.HE]}].

%
\bibitem{Gralla:2020srx}
S.~E.~Gralla, A.~Lupsasca, and D.~P.~Marrone,
Phys. Rev. D \href{https://doi.org/10.1103/PhysRevD.102.124004}{\textbf{102}, 124004 (2020)}
[arXiv:\href{https://arxiv.org/abs/2008.03879}{2008.03879 [gr-qc]}].

%
\bibitem{Darwin:1959}
C.~Darwin,
Proc. R. Soc. Lond. A \href{https://doi.org/10.1098/rspa.1959.0015}{\textbf{249}, 180 (1959)}.

%
\bibitem{Bozza:2001xd}
V.~Bozza, S.~Capozziello, G.~Iovane, and G.~Scarpetta,
Gen. Relativ. Gravit. \href{https://doi.org/10.1023/A:1012292927358}{\textbf{33}, 1535 (2001)}
[arXiv:\href{https://arxiv.org/abs/gr-qc/0102068}{gr-qc/0102068 [gr-qc]}].

%
\bibitem{Virbhadra:1999nm}
K.~S.~Virbhadra and G.~F.~R.~Ellis,
Phys. Rev. D \href{https://doi.org/10.1103/PhysRevD.62.084003}{\textbf{62}, 084003 (2000)}
[arXiv:\href{https://arxiv.org/abs/astro-ph/9904193}{astro-ph/9904193 [astro-ph]}].

%
\bibitem{Perlick:2004}
V.~Perlick,
Living Rev. Relativity \href{https://doi.org/10.12942/lrr-2004-9}{\textbf{7}, 9 (2004)}.

%
\bibitem{Bozza:2002zj}
V.~Bozza,
Phys. Rev. D \href{https://doi.org/10.1103/PhysRevD.66.103001}{\textbf{66}, 103001 (2002)}
[arXiv:\href{https://arxiv.org/abs/gr-qc/0208075}{gr-qc/0208075}].

%
\bibitem{Bozza:2010xqn}
V.~Bozza,
Gen. Relativ. Gravit. \href{https://doi.org/10.1007/s10714-010-0988-2}{\textbf{42}, 2269 (2010)}
[arXiv:\href{https://arxiv.org/abs/0911.2187}{0911.2187 [gr-qc]}].

%
\bibitem{Claudel:2000yi}
C.~M.~Claudel, K.~S.~Virbhadra, and G.~F.~R.~Ellis,
J. Math. Phys. (N.Y.) \href{https://doi.org/10.1063/1.1308507}{\textbf{42}, 818 (2001)}
[arXiv:\href{https://arxiv.org/abs/gr-qc/0005050}{gr-qc/0005050 [gr-qc]}].

%
\bibitem{Cvetic:2016bxi}
M.~Cvetic, G.~W.~Gibbons, and C.~N.~Pope,
Phys. Rev. D \href{https://doi.org/10.1103/PhysRevD.94.106005}{\textbf{94}, 106005 (2016)}
[arXiv:\href{https://arxiv.org/abs/1608.02202}{1608.02202 [gr-qc]}].

%
\bibitem{Kudo:2022ewn}
R.~Kudo and H.~Asada,
Phys. Rev. D \href{https://doi.org/10.1103/PhysRevD.105.084014}{\textbf{105}, 084014 (2022)}
[arXiv:\href{https://arxiv.org/abs/2201.01946}{2201.01946 [gr-qc]}].

%
\bibitem{Hod:2017zpi}
S.~Hod,
Phys. Lett. B \href{https://doi.org/10.1016/j.physletb.2017.11.021}{\textbf{776}, 1 (2018)}
[arXiv:\href{https://arxiv.org/abs/1710.00836}{1710.00836 [gr-qc]}].

%
\bibitem{Cunha:2017qtt}
P.~V.~P.~Cunha, E.~Berti, and C.~A.~R.~Herdeiro,
Phys. Rev. Lett. \href{https://doi.org/10.1103/PhysRevLett.119.251102}{\textbf{119}, 251102 (2017)}
[arXiv:\href{https://arxiv.org/abs/1708.04211}{1708.04211 [gr-qc]}].

%
\bibitem{Tsukamoto:2024pid}
N.~Tsukamoto,
Eur. Phys. J. C \href{https://doi.org/10.1140/epjc/s10052-024-13696-4}{\textbf{84}, 1325 (2024)}
[arXiv:\href{https://arxiv.org/abs/2401.07846}{2401.07846 [gr-qc]}].

%
\bibitem{Zhang:2024sgs}
J.~Zhang and Y.~Xie,
Phys. Rev. D \href{https://doi.org/10.1103/PhysRevD.109.043032}{\textbf{109}, 043032 (2024)}.

\bibitem{Tsukamoto:2020iez}
N.~Tsukamoto,
Phys. Rev. D \href{https://doi.org/10.1103/PhysRevD.102.104029}{\textbf{102}, 104029 (2020)}
[arXiv:\href{https://arxiv.org/abs/2008.12244}{2008.12244 [gr-qc]}].

%
\bibitem{Tsukamoto:2025hbz}
N.~Tsukamoto,
Phys. Rev. D \href{https://doi.org/10.1103/5yth-yfxt}{\textbf{113}, 104016 (2026)}
[arXiv:\href{https://arxiv.org/abs/2512.01688}{2512.01688 [gr-qc]}].

%
\bibitem{Chiba:2017nml}
T.~Chiba and M.~Kimura,
Prog. Theor. Exp. Phys. \href{https://doi.org/10.1093/ptep/ptx037}{\textbf{2017}, 043E01 (2017)}
[arXiv:\href{https://arxiv.org/abs/1701.04910}{1701.04910 [gr-qc]}].

%
\bibitem{Damour:2007ap}
T.~Damour and S.~N.~Solodukhin,
Phys. Rev. D \href{https://doi.org/10.1103/PhysRevD.76.024016}{\textbf{76}, 024016 (2007)}
[arXiv:\href{https://arxiv.org/abs/0704.2667}{0704.2667 [gr-qc]}].

%
\bibitem{Tsukamoto:2020uay}
N.~Tsukamoto,
Phys. Rev. D \href{https://doi.org/10.1103/PhysRevD.101.104021}{\textbf{101}, 104021 (2020); \textbf{106}, 049901(E) (2022)}
[arXiv:\href{https://arxiv.org/abs/2004.00822}{2004.00822 [gr-qc]}].

%
\bibitem{Fu:2021fxn}
Q.~M.~Fu and X.~Zhang,
Phys. Rev. D \href{https://doi.org/10.1103/PhysRevD.105.064020}{\textbf{105}, 064020 (2022)}
[arXiv:\href{https://arxiv.org/abs/2111.07223}{2111.07223 [gr-qc]}].

%
\bibitem{Patil:2016oav}
M.~Patil, P.~Mishra, and D.~Narasimha,
Phys. Rev. D \href{https://doi.org/10.1103/PhysRevD.95.024026}{\textbf{95}, 024026 (2017)}
[arXiv:\href{https://arxiv.org/abs/1610.04863}{1610.04863 [gr-qc]}].

%
\bibitem{Sasaki:2025web}
T.~Sasaki,
Phys. Rev. D \href{https://doi.org/10.1103/7r5r-nlcj}{\textbf{112}, 024072 (2025)}
[arXiv:\href{https://arxiv.org/abs/2504.00355}{2504.00355 [gr-qc]}].

%
\bibitem{Eiroa:2002mk}
E.~F.~Eiroa, G.~E.~Romero, and D.~F.~Torres,
Phys. Rev. D \href{https://doi.org/10.1103/PhysRevD.66.024010}{\textbf{66}, 024010 (2002)}
[arXiv:\href{https://arxiv.org/abs/gr-qc/0203049}{gr-qc/0203049}].

%
\bibitem{Tsukamoto:2016jzh}
N.~Tsukamoto,
Phys. Rev. D \href{https://doi.org/10.1103/PhysRevD.95.064035}{\textbf{95}, 064035 (2017)}
[arXiv:\href{https://arxiv.org/abs/1612.08251}{1612.08251 [gr-qc]}].

%
\bibitem{Igata:2025taz}
T.~Igata,
Phys. Rev. D \href{https://journals.aps.org/prd/abstract/10.1103/55vp-97gp}{\textbf{113}, 044042 (2026)}
[arXiv:\href{https://arxiv.org/abs/2503.02320}{2503.02320 [gr-qc]}].

%
\bibitem{Stefanov:2010xz}
I.~Z.~Stefanov, S.~S.~Yazadjiev, and G.~G.~Gyulchev,
Phys. Rev. Lett. \href{https://doi.org/10.1103/PhysRevLett.104.251103}{\textbf{104}, 251103 (2010)}
[arXiv:\href{https://arxiv.org/abs/1003.1609}{1003.1609 [gr-qc]}].

%
\bibitem{Raffaelli:2014ola}
B.~Raffaelli,
Gen. Relativ. Gravit. \href{https://doi.org/10.1007/s10714-016-2016-7}{\textbf{48}, 16 (2016)}
[arXiv:\href{https://arxiv.org/abs/1412.7333}{1412.7333 [gr-qc]}].

%
\bibitem{Igata:2026hzb}
T.~Igata and Y.~Takamori,
arXiv:\href{https://arxiv.org/abs/2603.09946}{2603.09946 [gr-qc]}.

%
\bibitem{Igata:2025plb}
T.~Igata,
Phys. Rev. D \href{https://doi.org/10.1103/PhysRevD.113.024036}{\textbf{113}, 024036 (2026)}
[arXiv:\href{https://arxiv.org/abs/2504.07906}{2504.07906 [gr-qc]}].

%
\bibitem{Igata:2025hpy}
T.~Igata,
arXiv:\href{https://arxiv.org/abs/2505.01848}{2505.01848 [gr-qc]}.

%
\bibitem{Wald:1984}
R.~M.~Wald,
\textit{General Relativity} (University of Chicago Press, Chicago, 1984).

%
\bibitem{Misner:1964je}
C.~W.~Misner and D.~H.~Sharp,
Phys. Rev. \href{https://doi.org/10.1103/PhysRev.136.B571}{\textbf{136}, B571 (1964)}.

%
\bibitem{Hayward:1994bu}
S.~A.~Hayward,
Phys. Rev. D \href{https://doi.org/10.1103/PhysRevD.53.1938}{\textbf{53}, 1938 (1996)}
[arXiv:\href{https://arxiv.org/abs/gr-qc/9408002}{gr-qc/9408002}].

%
\bibitem{Kinoshita:2024wyr}
S.~Kinoshita,
Phys. Rev. D \href{https://doi.org/10.1103/PhysRevD.110.044056}{\textbf{110}, 044056 (2024)}
[arXiv:\href{https://arxiv.org/abs/2402.16484}{2402.16484}].

%
\bibitem{Shaikh:2019itn}
R.~Shaikh, P.~Banerjee, S.~Paul, and T.~Sarkar,
Phys. Rev. D \href{https://doi.org/10.1103/PhysRevD.99.104040}{\textbf{99}, 104040 (2019)}
[arXiv:\href{https://arxiv.org/abs/1903.08211}{1903.08211 [gr-qc]}].

%
\bibitem{Tsukamoto:2021fsz}
N.~Tsukamoto,
Phys. Rev. D \href{https://doi.org/10.1103/PhysRevD.104.124016}{\textbf{104}, 124016 (2021)}
[arXiv:\href{https://arxiv.org/abs/2107.07146}{2107.07146 [gr-qc]}].

%
\bibitem{Hayward:2005gi}
S.~A.~Hayward,
Phys. Rev. Lett. \href{https://doi.org/10.1103/PhysRevLett.96.031103}{\textbf{96}, 031103 (2006)}
[arXiv:\href{https://arxiv.org/abs/gr-qc/0506126}{gr-qc/0506126 [gr-qc]}].

%
\bibitem{Bardeen:1968}
J.~M.~Bardeen,
in \textit{Proceedings of the 5th International Conference on Gravitation and the Theory of Relativity (GR5)}, 
edited by V.~A.~Fock \textit{et al.} (Tbilisi University Press, Tbilisi, 1968), p. 174.

%
\bibitem{Ayon-Beato:2000mjt}
E.~Ayon-Beato and A.~Garcia,
Phys. Lett. B \href{https://doi.org/10.1016/S0370-2693(00)01125-4}{\textbf{493}, 149 (2000)}
[arXiv:\href{https://arxiv.org/abs/gr-qc/0009077}{gr-qc/0009077 [gr-qc]}].

%
\bibitem{Eiroa:2010wm}
E.~F.~Eiroa and C.~M.~Sendra,
Classical Quantum Gravity \href{https://doi.org/10.1088/0264-9381/28/8/085008}{\textbf{28}, 085008 (2011)}
[arXiv:\href{https://arxiv.org/abs/1011.2455}{1011.2455 [gr-qc]}].

%
\bibitem{Lemos:2008cv}
J.~P.~S.~Lemos and O.~B.~Zaslavskii,
Phys. Rev. D \href{https://doi.org/10.1103/PhysRevD.78.024040}{\textbf{78}, 024040 (2008)}
[arXiv:\href{https://arxiv.org/abs/0806.0845}{0806.0845 [gr-qc]}].

%
\bibitem{Berti:2009kk}
E.~Berti, V.~Cardoso, and A.~O.~Starinets,
Classical Quantum Gravity \href{https://doi.org/10.1088/0264-9381/26/16/163001}{\textbf{26}, 163001 (2009)}
[arXiv:\href{https://arxiv.org/abs/0905.2975}{0905.2975 [gr-qc]}].

%
\bibitem{Konoplya:2011qq}
R.~A.~Konoplya and A.~Zhidenko,
Rev. Mod. Phys. \href{https://doi.org/10.1103/RevModPhys.83.793}{\textbf{83}, 793 (2011)}
[arXiv:\href{https://arxiv.org/abs/1102.4014}{1102.4014 [gr-qc]}].

%
\bibitem{Ferrari:1984zz}
V.~Ferrari and B.~Mashhoon,
Phys. Rev. D \href{https://doi.org/10.1103/PhysRevD.30.295}{\textbf{30}, 295 (1984)}.

%
\bibitem{Cardoso:2008bp}
V.~Cardoso, A.~S.~Miranda, E.~Berti, H.~Witek, and V.~T.~Zanchin,
Phys. Rev. D \href{https://doi.org/10.1103/PhysRevD.79.064016}{\textbf{79}, 064016 (2009)}
[arXiv:\href{https://arxiv.org/abs/0812.1806}{0812.1806 [hep-th]}].

%
\bibitem{Yang:2012pj}
H.~Yang, F.~Zhang, A.~Zimmerman, D.~A.~Nichols, E.~Berti, and Y.~Chen,
Phys. Rev. D \href{https://doi.org/10.1103/PhysRevD.87.041502}{\textbf{87}, 041502 (2013)}
[arXiv:\href{https://arxiv.org/abs/1212.3271}{1212.3271 [gr-qc]}].

%
\bibitem{Yang:2012he}
H.~Yang, D.~A.~Nichols, F.~Zhang, A.~Zimmerman, Z.~Zhang, and Y.~Chen,
Phys. Rev. D \href{https://doi.org/10.1103/PhysRevD.86.104006}{\textbf{86}, 104006 (2012)}
[arXiv:\href{https://arxiv.org/abs/1207.4253}{1207.4253 [gr-qc]}].

%
\bibitem{Iyer:1986np}
S.~Iyer and C.~M.~Will,
Phys. Rev. D \href{https://doi.org/10.1103/PhysRevD.35.3621}{\textbf{35}, 3621 (1987)}.

%
\bibitem{Hatsuda:2021gtn}
Y.~Hatsuda and M.~Kimura,
Universe \href{https://doi.org/10.3390/universe7120476}{\textbf{7}, 476 (2021)}
[arXiv:\href{https://arxiv.org/abs/2111.15197}{2111.15197 [gr-qc]}].

%
\bibitem{Konoplya:2017wot}
R.~A.~Konoplya and Z.~Stuchl\'{\i}k,
Phys. Lett. B \href{https://doi.org/10.1016/j.physletb.2017.06.015}{\textbf{771}, 597 (2017)}
[arXiv:\href{https://arxiv.org/abs/1705.05928}{1705.05928 [gr-qc]}].

%
\bibitem{Guo:2021enm}
G.~Guo, P.~Wang, H.~Wu, and H.~Yang,
J. High Energy Phys. \href{https://doi.org/10.1007/JHEP06(2022)060}{06 (2022) 060}
[arXiv:\href{https://arxiv.org/abs/2112.14133}{2112.14133 [gr-qc]}].

%
\bibitem{Guo:2023mda}
G.~Guo, P.~Wang, H.~Wu, and H.~Yang,
J. High Energy Phys. \href{https://doi.org/10.1007/JHEP10(2023)076}{10 (2023) 076}
[arXiv:\href{https://arxiv.org/abs/2307.12210}{2307.12210 [gr-qc]}].

%
\bibitem{Teo:2003ltt}
E.~Teo,
Gen. Relativ. Gravit. \href{https://doi.org/10.1023/A:1026286607562}{\textbf{35}, 1909 (2003)}.


\end{thebibliography}
\end{document}